\documentclass[twoside]{ilcws07}
\usepackage[latin1]{inputenc}
\usepackage[dvips]{graphicx,epsfig,color}
\usepackage{wrapfig,rotating}
\usepackage{amssymb,amsmath,array}

\begin{document}
\title{
R\&D STATUS OF ATF2 IP BEAM SIZE MONITOR (SHINTAKE MONITOR)} 
\author{T.~Suehara$^1$, H.~Yoda$^1$, M.~Oroku$^1$, T.~Yamanaka$^1$, S.~Komamiya$^1$,\\
T.~Sanuki$^2$, T.~Tauchi$^3$, Y.~Honda$^4$ and T.~Kume$^5$ 
\vspace{.3cm}\\
1- The University of Tokyo - Department of Physics \\
7-3-1 Hongo, Bunkyo, Tokyo - Japan
\vspace{.1cm}\\
2- Tohoku University - Department of Physics \\
6-3 Aoba, Aramaki, Aoba, Sendai, Miyagi - Japan
\vspace{.1cm}\\
3- KEK - Institute of Particle and Nuclear Studies \\
4- KEK - Accelerator Laboratory \\
5- KEK - Applied Research Laboratory \\
1-1 Oho, Tsukuba, Ibaraki - Japan
}

\maketitle

\begin{abstract}
Shintake monitor\cite{suehara_taikan:shintake} is a nanometer-scale electron beam size monitor.
It probes a electron beam by an interference fringe pattern formed by split laser beams.
Minimum measurable beam size by this method is less than 1/10 of laser wavelength.
In ATF2, Shintake monitor will be used for the IP beam size monitor to measure 37 nm (design) beam size.
Development status of the Shintake monitor, including fringe phase monitoring and stabilization,
gamma detector and collimators, is described. In addition, we discuss the beam size measurement
by Shintake monitor in ILC.
\end{abstract}

\section{Overview}
\subsection{Shintake Monitor}

\begin{wrapfigure}{r}{0.5\columnwidth}
\centerline{\includegraphics[width=0.45\columnwidth]{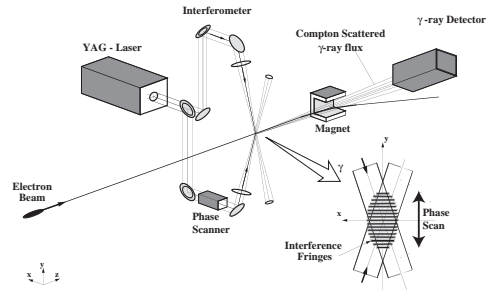}}
\caption{A schematic of Shintake monitor. Original figure from \cite{suehara_taikan:shintake_proc}, partly revised.}\label{suehara_taikan.fig.schematic}
\end{wrapfigure}

Figure \ref{suehara_taikan.fig.schematic} shows a schematic of Shintake monitor. A photon beam from a YAG Laser (2nd harmonics, 532 nm wavelength) is split
and go across the focal point of the electron beam line from the opposite direction to make an interference fringe pattern.
Photons in the fringe interact with the electron beam by inverse-Compton scattering process.

The fringe pattern behaves as a kind of a modulation of photon density, so number of scattered photons can be modulated
by scanning the phase of the laser fringe on the IP.
Depth of the modulation of Compton photon density $(N/N_0)$ depends on the electron beam size by,
\begin{equation}
	\frac{\Delta{}N}{N_0} = \exp\left(-\frac{\left(2k_0\sigma_y\right)^2}{2}\right)
	\label{suehara_taikan.eqn.moddepth}
\end{equation}
where $k_0$ is the laser wavenumber and $\sigma_y$ is the beam size of y axis.

We can obtain the electron beam size by measuring this modulation depth
with a gamma-ray monitor located downstream of the IP.

\subsection{Required Performance for ATF2}

ATF2 (Accelerator Test Facility 2)\cite{suehara_taikan:proposal} is a final focus test bench for ILC.
It has 2 major goals, which are achievement and maintenance of 37 nm beam size by ILC-like beam optics
and stabilization of beam position to nanometer level.

Shintake monitor is the key component to realize the first goal. It will be used for beam tuning as well as for
confirming the achievement of 37 nm beam size.
To meet the goal, 2 nm resolution will be required for the Shintake monitor.
By Equation (\ref{suehara_taikan.eqn.moddepth}),
2 nm measurement error for 37 nm beam size corresponds to 3 \% error of the modulation depth.
In the following sections, we focus on the key techniques to realize 3 \% resolution of the modulation depth. 

\section{Phase Control of the Laser Fringe}

\subsection{Required Fringe Stability}

\begin{wrapfigure}{r}{0.5\columnwidth}
\centerline{\includegraphics[width=0.45\columnwidth]{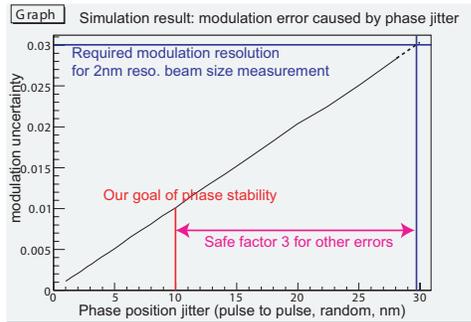}}
\caption{Effect of phase jitter on the modulation error.}\label{suehara_taikan.fig.phasejitter_modulation}
\end{wrapfigure}

Phase fluctuaion of the laser fringe on IP is one of the major error sources for the Shintake monitor.
We performed a simulation study to estimate the stability requirement for 3 \% modulation resolution. 

The simulation result is shown in Figure \ref{suehara_taikan.fig.phasejitter_modulation}. The graph shows that 3\% resolution of the modulation depth
requires 30 nm phase position stability. Considering other error factors, we should stabilize the fringe phase to 10 nm level.
Note that the simulated position error is pulse-to-pulse jitter, and the real phase fluctuation may include slower vibration or drift, but we think the safety factor of 3 should suppress the error caused by the simulation model.

\subsection{Phase Monitor \& Control}

To achieve 10 nm phase stabilization, active phase stabilization system was developed.
The phase monitor consists of a image sensor (Hamamatsu S9226)
and a microscope lens (Nikon CF IC EPI Plan 100 x A).
Split laser beams are guided into the microscope lens, forming an interference fringe pattern
captured by the image sensor located at the back of the lens.
The pixel size of the image sensor is 7.8 $\mu$m, which is much larger than the fringe pitch at IP, 
but fringe magnification effect by the microscope lens makes the fringe pitch broaden to 
the observable range (a few tens of $\mu$m, depends on beam entering angle to the lens and the distance
between the lens and the image sensor). 
An obtained spectrum by the image sensor is Fourier-transformed (online),
and the phase at the peak of the power spectrum is used as the detected phase.

For phase control, we installed an optical delay line with a piezo stage (PI P-752.21C)
to one of the split laser beam line. The stage has 0.2 nm resolution, and we use a 16 bit
VME DAC (Advanet Advme2706) to indicate the position. The resolution on DAC is about 0.05 nm
after the output voltage is reduced to 1/10 by resistor split.
The frequency of feedback control is 10 Hz, that is the repetition rate of the pulsed laser.
For feedback algorithm, software PID control is implemented.

\subsection{Result of Phase Stabilization}

Because the fringe phase at IP cannot be directly measured,
effect of the phase stabilization was evaluated by 2 image sensors.
We use 1 sensor for stabilization and monitor another sensor to obtain the stabilization effect.
The position of the sensors is selected symmetrically over IP
(See the slide\cite{url} for the geometry).

The stabilization result shows 0.034 radian (1.5 nm) stability in 1 minute window, and
0.133 radian (5.6 nm) stability in 10 minutes window
using continuous-wave low power laser.
Both meet 10 nm stability requirement of ATF2, but the stability is strongly depends on 
environmental conditions, including ground and air motion, temperature shift etc.

Stabilization study on pulsed laser is going on. Stability achieved up to now seems slightly worse,
but we can expect 10 nm stabilization should be possible.

\section{Collimators and Gamma Detectors}

\subsection{Beam Halo and Electron Collimator}

The major background of the gamma detector is photons emitted by 
the electron beam halo hitting the beam pipe. Distribution of ATF2 beam halo was measured
and the result was reported in \cite{suehara_taikan:nanobeam}.
Because the beam size at final focus magnets must be very large for strong focusing,
the beam tail must be cut upstream to prevent background photons.

In ATF2, BPMs of upstream optics work as the collimators, but the effect should be confirmed
by tracking simulation. We are preparing the tracking simulation now.

\subsection{Gamma Detector}

The proposed gamma detector for the Shintake monitor consists of several layers of CsI(Tl) scintillator.
Thickness of layers is 10 mm for forward 4 layers, and 300 mm for a rear single layer.
The average energy of Compton signal and background from beam pipe is much different,
and with the forward layers we can obtain the S/N ratio using this energy difference.
The result of simulation study is shown in \cite{suehara_taikan:gamma}.

Now we are going on detailed design and assembly of the detector.

\subsection{Gamma Collimator}

\begin{wraptable}{r}{0.6\columnwidth}
\small{
\centerline{
		\begin{tabular}{|c|c|c|}\hline
			Aperture & Signal / BG & Signal enhance \\
			angle from IP & acceptance & ratio \\ \hline\hline
			2.20 mrad & 95 \% / 5 \% & 19 \\\hline
			1.30 mrad & 80 \% / 1 \% & 80 \\ \hline
			0.83 mrad & 60 \% / 0.1 \% & 600 \\ \hline
		\end{tabular}}}
	\caption{Suppression ratio by gamma collimators.}
	\label{suehara_taikan.tbl.collimator}
\end{wraptable}	

To reduce background photons at the detector, we plan to install a gamma collimator
in front of the detector. The optimal collimator is cone-shaped, because signal photons are emitted from a point-source
at IP and strictly restricted to forward angle by Lorenz-boost kinematics.

The optimal radius of the collimator depends on S/N ratio.
The fraction of background passing the collimator is strongly suppressed when narrowing 
the radius of the collimator, while the fraction of signal passing is rather mildly decreased.
So, if the S/N ratio is very bad, narrower collimator is favored, while wider collimator is 
better when S/N ratio is not so bad. Suppression ratios by typical opening angles of the collimator
is shown in Table \ref{suehara_taikan.tbl.collimator}.

\section{Shintake Monitor for ILC}

As a sub-micron beam size monitor, Shintake monitor can be useful for ILC or other future colliders.
For using in ILC, several consideration should be needed.

\begin{itemize}
	\item Because the beam energy of ILC is much larger than ATF2, the cross section of Compton scattering
	is lower, about 1/10 of ATF2. We need more laser beam energy or stronger laser focusing to obtain 
	statistics enough.
	\item As the peak energy of the Compton scattering photons is almost the same as the beam energy, energy separation
	of signal and pipe-scattered background is not realistic. In addition, the energy of synchrotron radiation 
	photons from focusing magnets is also larger in ILC, which should be cut by some kind of shields in front
	of the gamma detector.
	\item IP beam size of ILC is about 5 nm. For measuring 5 nm beam size, wavelength of the laser beam should
	be minimized. Within commercially available lasers, 193 nm excimer laser may have the shortest wavelength
	for high power pulsed beam.
	Using a 193 nm laser, 5 nm measurement is not impossible. Assuming the same resolution of the modulation depth 
	measurement as ATF2 goal, the resolution on beam size is about $\pm$1 nm (20 \%).
\end{itemize}

\section{Summary and Outlook}

Shintake monitor will be installed as an IP-BSM in ATF2 to measure 37 nm electron beam size.
10 nm stability of the laser fringe is necessary to achieve less than 5 \% error on beam size measurement,
and implementation of 10 nm level fringe stabilization is almost finished.
Study and design of the gamma detector and collimators are going on.
We plan to install the Shintake monitor in ATF2 IP region in early 2008 with the fringe stabilization system.
Beam test of the gamma detector will be performed in this autumn and winter.
By beginning of ATF2 commissioning run planned in end of 2008, the Shintake monitor will be ready for 37 nm 
beam size measurement (after some adjustment and tuning using the electron beams).


\begin{footnotesize}



%

\end{footnotesize}


\end{document}